# *MIK2* is a candidate gene of the S-locus for sporophytic self-incompatibility (SSI) in chicory (*Cichorium intybus,* Asteraceae)

Fabio Palumbo[†], Samela Draga[†], Gabriele Magon[†], Giovanni Gabelli, Alessandro Vannozzi, Silvia Farinati, Francesco Scariolo, Margherita Lucchin, and Gianni Barcaccia*

Laboratory of Plant Genetics and Breeding, Department of Agronomy, Food, Natural Resources, Animals and Environment (DAFNAE), University of Padua, Padua, Italy

[†]These authors have contributed equally to this work

* Correspondence:
Gianni Barcaccia
Gianni.barcaccia@unipd.it

Keywords: *MDIS1 INTERACTING RECEPTOR LIKE KINASE 2*, pollen–stigma recognition, AT4G08850, endive, reproductive systems

## Abstract

The *Cichorium* genus offers a unique opportunity to study the sporophytic self-incompatibility (SSI) system, being composed of species characterized by highly efficient SI (*C. intybus*) and complete self-compatibility (*C. endivia*). The chicory genome was used to map 7 previously identified SSI locus-associated markers. The region containing the S-locus was restricted to an ~4 M bp window on chromosome 5. Among the genes predicted in this region, *MDIS1 INTERACTING RECEPTOR LIKE KINASE 2 (MIK2)* was promising as a candidate for SSI. Its ortholog in Arabidopsis is involved in pollen–stigma recognition reactions, and its protein structure is similar to that of S-receptor kinase (SRK), a key component of the SSI in the *Brassica* genus. The sequencing of *MIK2* in chicory and endive accessions revealed two contrasting scenarios. In *C. endivia*, MIK2 was fully conserved even comparing different botanical varieties (smooth and curly). In *C. intybus*, 387 SNPs and 3 INDELs were identified when comparing accessions of different biotypes from the same botanical variety (radicchio). The SNP distribution throughout the gene was uneven, with hypervariable domains preferentially localized in the LRR-rich extracellular region, putatively identified as the receptor domain. The gene was hypothesized to be under positive selection, as the nonsynonymous mutations were more than double the synonymous ones (dN / dS = 2.17). An analogous situation was observed analyzing the first 500 bp of the *MIK2* promoter: no SNPs were observed among the endive samples, whereas 44 SNPs and 6 INDELs were detected among the chicory samples. Further analyses are needed to confirm the role of MIK2 in SSI and to demonstrate whether the 23 species-specific nonsynonymous SNPs in the CDS and/or the species-specific 10 bp-INDEL found in a CCAAT box region of the promoter are responsible for the contrasting sexual behaviors of the two species.

## 1 Introduction

Self-incompatibility (SI) is a peculiar evolutionary strategy aimed at producing and preserving high levels of genetic variability within a species, which prevents self-fertilization and thus inbreeding depression (de Nettancourt, 2001). SI is a common feature in flowering plants, occurring in approximately 40% of angiosperm species (Saumitou-Laprade et al., 2017). It results in the total or partial lack of germination of the pollen grain or development of the pollen tube due to specific interactions between pollen grain and stigma surface or transmitting tissue of the style (Ahmad et al., 2022). As far as is known, SI is prevalently controlled by a single multiallelic locus, the S-locus (Brom et al., 2020), and the combinations of the different allelic variants composing the locus define the S-haplotypes. SI occurs when the same S-haplotype is expressed by both male (pollen) and female (stigma or style) interacting tissues (Takayama and Isogai, 2005). There are two main types of SI: gametophytic SI (GSI) and sporophytic SI (SSI).

In GSI, incompatible mating occurs when the S-locus carried by the haploid pollen (male gametophyte) matches either of the S loci present in the diploid style (female sporophyte). As a result, incompatible pollen germinates successfully on the stigma surface, penetrates the stigma, and grows into the style, but pollen tube growth is arrested through the transmitting tract toward the ovary (Broz and Bedinger, 2021). This kind of SI has been described in several plant families, including Fabaceae, Poaceae, Rosaceae, Plantaginaceae and Solanaceae (Watanabe et al., 2012).

In contrast, SSI is common in the Brassicaceae, Convolvulaceae, Oleaceae and Asteraceae families (Alagna et al., 2019; Price et al., 2022) and is the result of a more complex mechanism. In this case, the behavior of the pollen is determined by the diploid S genotype of the pollen-producing plant (male sporophyte) so that diploid sporophytic expression of the S-locus allows dominance interactions to occur between male and female tissues (Novikova et al., 2023). Moreover, unlike GSI, pollen tube growth in incompatible mating is immediately arrested on the surface of the stigma. The molecular mechanisms underlying SSI have been extensively deepened in the Brassicaceae family (for a comprehensive review, see (Abhinandan et al., 2022)), whereas in the Asteraceae family, the SSI systems have been poorly investigated and are even less well understood. In this family, the *Cichorium* genus offers a unique opportunity, being composed of two main groups: one characterized by a strong SSI behavior (*C. intybus*, *C. spinosum*, and *C. bottae*) and the other group containing all self-compatible species (*C. endivia*, *C. pumilum*, and C. *calvum*) (Lucchin et al., 2008). Most interestingly, a cladistic analysis conducted by Kiers et al. and based on the combined use of restriction fragment length polymorphisms (AFLP), trnL-trnF and ITS1/2 sequences indicated that the contrasting sexual behavior of these two groups seems to be strictly related to the phylogeny of the genus (Kiers et al., 1999).

In *C. intybus*, SSI was first demonstrated by analyzing different combinations of crosses between Witloof chicory inbred lines (Eenink, 1981). SSI was also confirmed by crossing wild-type chicory plants with accessions of the Italian biotype Rosso di Chioggia (Varotto et al., 1995). It was observed that SSI in chicory induced a quick rejection process, which, after a few minutes, provokes the inhibition of pollen hydration or germination (Barcaccia et al., 2016). From a molecular point of view, Gonthier et al. (Gonthier et al., 2013) assigned the genetic determination of SI to a single S-locus located in LG2. Beyond this, the lack of a good-quality genome assembly has made the identification of putative S-loci extremely challenging. The recent release of the endive and chicory genomes (Fan et al., 2022), together with the availability of several S-locus-associated molecular markers developed in the last decade (Gonthier et al., 2013), represent a turning point in the study of SSI. Since SI hampers the production of inbred lines in several Italian biotypes of chicory, the identification of the S-locus is of pivotal importance for breeding programs.

## 2    Materials and Methods

### 2.1    *In silico* identification of SSI candidate genes

The newly released reference genomes of chicory (*Cichorium intybus*) and endive (*Cichorium endivia*) were first retrieved from NCBI (JAKNSD000000000 and JAKOPN000000000, respectively) (Fan et al., 2022)).

Two SSR markers, namely, sw2H09.2 and B131, and five AFLP-derived markers, all located within the same linkage group (LG2), were selected because of their association with the SSI locus in *C. intybus* (Cadalen et al., 2010; Gonthier et al., 2013). The main features are reported in **Table 1**.

**Table 1.** Molecular markers cosegregating with the S-locus in *C. intybus* (Gonthier et al., 2013). sw2H09.2[1] and B131[1] are microsatellite (SSR) regions, while the remaining are AFLP-derived markers. For each marker, the name, reference, putative distance from the SSI locus and primer sequence are indicated.

| Marker name | NCBI accession | Distance from SSI locus[2] | Forward Primer | Reverse Primer |
|---|---|---|---|---|
| sw2H09.2[1] | n.a. | ~9 cM - downstream | GTGCCGGTCTTCAGGTTACA | CGCCTACCGATTACGATTGA |
| B131[1] | n.a. | ~45 cM - downstream | CCGCTCTCTCATCACTCCTC | GCTCGAAAATCGGCTACAAC |
| TACG294[2] | GF112133.1 | < 1 cM - upstream | TCCCTTCAATGAGTCGATGT | TGGAATAAATTCAGCCATCCT |
| TACG293[2] | GF112134.1 | < 1 cM - upstream | TCCCTTCAATGAGTCGATGT | TGGAATAAATTCAGCCATCCT |
| AACC134[2] | GF112135.1 | < 1 cM - downstream | ACCCCAAATTTCAGGTTTC | CAAAATAGTTCAGGTGACTTACGC |
| TTAA440[2] | GF112136.1 | < 1 cM - downstream | CAATGCGTGCCTTTTGTATG | CAACCAAATCATCTCTTCTCTCTC |
| GGATT128[2] | GF112137.1 | < 1 cM - upstream | CAAGTCAGCCTCCCAAACAT | ATTCAGGTGCAGGAGGACAT |

[1](Cadalen et al., 2010)

[2](Gonthier et al., 2013)

These seven markers were mapped against the *C. intybus* genome to identify the SSI locus-carrying chromosome corresponding to LG2 of Cadalen et al. (Cadalen et al., 2010) and, therefore, to narrow down the region containing the putative SI locus. Similarly, the same markers were also mapped against the *C. endivia* genome for a chromosome-level comparison between the two *Cichorium* species. Mapping analyses were performed using Geneious Prime 2022.2.1 software (https://www.geneious.com).

All the predicted amino acid sequences included within the chromosome window carrying the SSI locus in chicory were annotated locally via BLASTp against the proteomes of *Arabidopsis thaliana* (Araport 11 (Cheng et al., 2017)) and *Lactuca sativa* (Lsat Salinas v11 (Reyes-chin-wo et al., 2017), as a model species of the Asteraceae family.

The candidate gene selected after the abovementioned analyses (KAI3736590.1) was further investigated in order i) to identify any possible protein domain and/or functional site using PROSITE (Sigrist et al., 2010) at Expasy (Duvaud et al., 2021) and ii) to predict the topology of both alpha-helical and beta-barrel transmembrane domains by means of DeepTMHMM (https://dtu.biolib.com/DeepTMHMM). The orthologs of the SSI-related candidate gene in chicory were also searched in endive.

## 2.2  *In vitro* testing of the putative candidate determinant of SSI

Fifteen samples from *C. intybus* and *C. endivia*, representing the main cultivated biotypes traditionally available in the Veneto region (Italy), were used to evaluate the polymorphism rate of the putative SSI determinant. For *C. intybus*, 12 local biotypes of Radicchio, all belonging to the botanical var. *foliosum*, (i.e., 4 Rosso di Chioggia, 2 Variegato di Castelfranco, 2 Precoce di Treviso, 1 Tardivo di Treviso, 1 Verona Semilungo, 1 Variegato di Chioggia and 1 Rosa) were retrieved from the market. For *C. endivia,* we collected 2 samples from the botanical var. *latifolium* (smooth endive) and 1 sample from the botanical var. *crispum* (curly endive).

Genomic DNA (gDNA) was isolated from 100 mg of fresh leaves using a DNAeasy Plant Mini Kit (Qiagen, Valencia, CA, USA) following the procedure provided by the manufacturer. The quality and quantity of the gDNA were assessed by agarose gel electrophoresis (1% agarose/1× TAE gel containing 1× SYBR Safe DNA Stain, Life Technologies, Carlsbad, CA, USA) and a NanoDrop 2000c UV–Vis spectrophotometer (Thermo Fisher Scientific Inc., Pittsburgh, PA, USA), respectively.

Five primer pairs were designed to span the entire sequence of the candidate gene (3193 bp) previously selected. Primers were drawn using Geneious Prime 2022.2.1 in the most conserved regions between *C. intybus* and *C. endivia* to allow successful amplification of samples from both species (**Table 2**). Similarly, a further primer pair was synthesized to amplify the first 500 bp upstream of the start codon, allowing for a comparison between the putative promoter regions of the two species.

**Table 2.** List of primer pairs used for the amplification of two genes in *C. intybus* and *C. endivia*. KAI3736590.1 was selected as one of the possible determinants of the S-locus, and KAI3736550.1 was chosen from the same chromosomal region for a comparison between the polymorphism rates of the two loci. Due to the length of the two genes, five and two primer pairs (for the amplification of as many overlapping regions) were designed to cover their entire sequences. For the candidate gene, we also designed a primer pair to amplify the first 500 bp of the promoter region. For each primer pair, the sequence, temperature of melting (Tm), and distance from the start codon (ATG) are indicated. The primers used for Sanger sequencing are highlighted in bold.

| Gene | Region name | Forward | Tm | Distance From ATG | Reverse | Tm | Distance From ATG | Amplified region length |
|---|---|---|---|---|---|---|---|---|
| | Prom500 | **CCACTCGATSTTCTCATGTAC** | 62.8 | -561 | **AGAAGGGCAGTGAATTCATC** | 61.2 | 119 | 680 |
| | Part_1 | **CCACATAAAACTTCCTCCAATTTG** | 61.6 | -82 | CAGGGATGGGACCAGAAAGG | 65.1 | 976 | 1058 |
| KAI3736590.1 | Part_2 | GGTAACTTGACCAACCTCAGG | 63.4 | 624 | **GCCCTTCCAAATTATTCATTGAC** | 61.4 | 1684 | 1060 |
| (3193 bp) | Part_3 | TGGTGTGTACCCTAGCCTCA | 65.4 | 1340 | **CACCTTTCCGCATCTCCCTT** | 65.3 | 2349 | 1009 |
| | Part_4 | **CTTGGAAGGTCCCATTCCCA** | 64.9 | 1961 | TATAGGGGGTGAGCAGTCGT | 65.2 | 2670 | 709 |
| | Part_5 | AAGGGAGATGCGGAAAGGTG | 65.3 | 2329 | **GAGAACCACATAATTTGACAGCC** | 62.8 | 3280 | 951 |
| KAI3736550.1 | Part_1 | **CCACTATGCTCTGTTGTTATACT** | 60.9 | -62 | CGTTTCGGCCGGAAAAGATC | 64.6 | 796 | 858 |
| (1491 bp) | Part_2 | GGCGATTTCGGTTTGGCTTT | 65.0 | 618 | **TATCGTGTCAAATATGCACTGC** | 62.1 | 1518 | 900 |

In addition, from the same chromosomal region containing the SSI locus, we selected a second gene (KAI3736550.1) whose ortholog in Arabidopsis does not appear to be involved in pollen–stigma recognition. This gene was chosen as a control for a comparison between its polymorphism rate and that of the candidate gene. Additionally, in this case, two primer pairs were designed to cover its entire sequence and to allow successful amplification of samples from both *C. intybus* and *C. endivia* (**Table 2**).

PCRs were performed using ~30 ng of gDNA as a template, 10 µL of MangoMix (Bioline, London, United Kingdom), 2 µL of each primer (10 mM) and sterile water to a final volume of 20 µL. A Veriti 96-Well Thermal Cycler (Applied Biosystems, Carlsbad, CA) was used to carry out the amplifications by setting the following conditions: initial denaturation at 95 °C for 5 min, followed by 35 cycles at 95 °C for 30 s, 59 °C for 30 s, and 72 °C for 90 s, and a final extension of 10 min at 72 °C. The quality of the PCR amplicons was assessed on a 1.5% (w/v) agarose gel stained with 1× SYBR Safe DNA Gel Stain (Life Technologies). Amplicons were purified with ExoSap-IT (Applied Biosystems), sequenced through Sanger sequencing, analyzed and manually curated in Geneious 2022.2.1, assembled and finally deposited in GenBank (accession nos. OQ781894-OQ781923).

## 3 Results and Discussion

### 3.1 Identification of *MIK2* in *Cichorium* species

The mapping of the seven markers associated with the SSI locus (Gonthier et al., 2013) on the newly released reference genome of chicory allowed us to identify the correspondence between LG2 (Cadalen et al., 2010; Gonthier et al., 2013) and chromosome 5 (JAKNSD000000000 (Fan et al., 2022)). Most importantly, this allowed us to narrow down the region containing the SSI locus to a window of ~4 M bases, between ~167,000 bp and ~4,235,000 bp (**Figure 1A**).

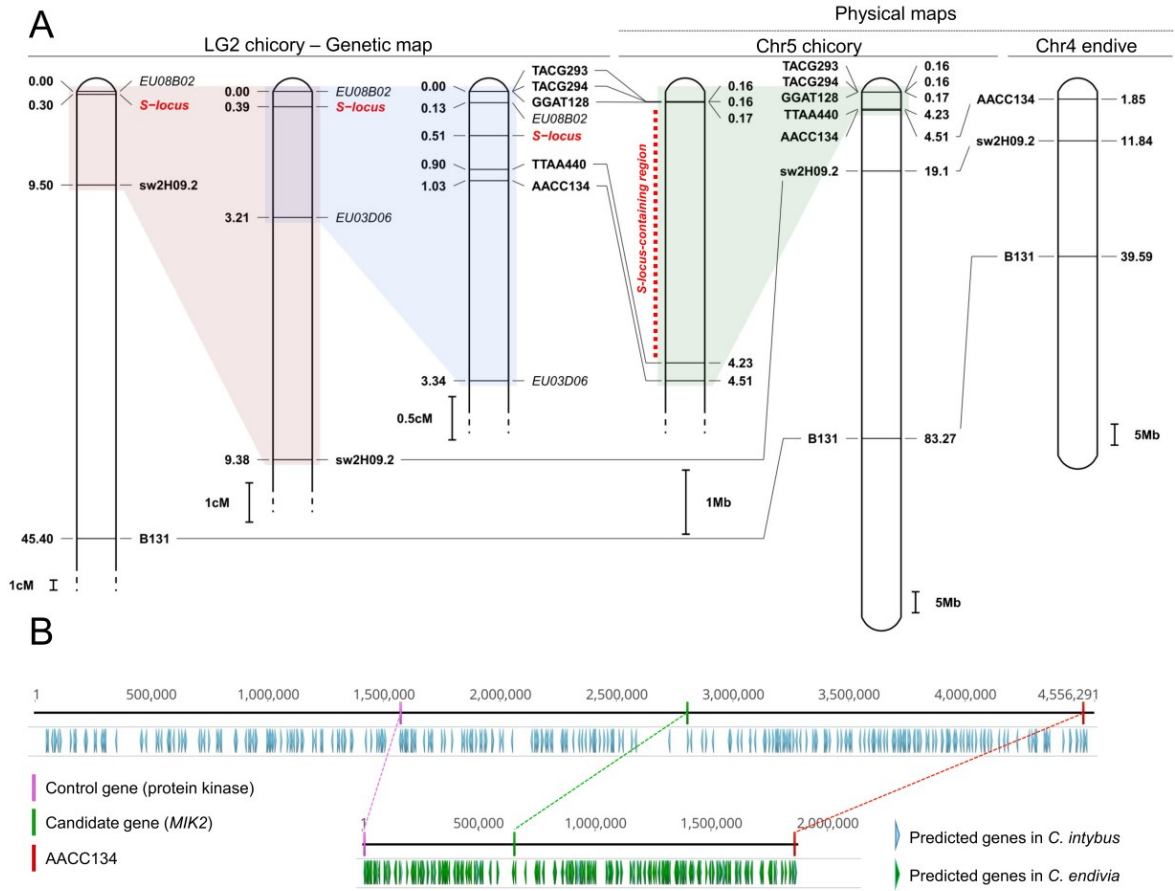

**Figure 1**. Narrowing down the chromosomal window containing the SSI locus and identification of a candidate gene. **(A)** Correspondence between LG2 of *Cichorium intybus* (Gonthier et al., 2013), chromosome 5 of *C. intybus* and chromosome 4 of *C. endivia* (Fan et al., 2022). Seven molecular markers (B131 and swH09.2, TACG2942, TACG2932, AACC1342, TTAA4402, GGATT1282, in bold), originally identified by Gonthier et al. because of their association with the S-locus (in red), were mapped against the genome of chicory (JAKNSD000000000), and the SSI-locus was localized in a window of 4 M bases on chromosome 5 (red dotted line). Three out of seven markers were also successfully mapped against the endive genome (JAKOPN000000000), allowing us to identify the correspondence between chromosome 4 and chromosome 5 of chicory. Markers in italics represent other molecular markers identified by Gonthier et al. and associated with the SSI locus whose sequences are not available in NCBI. **(B)** Comparison between the region of ~4 M bp located on the peripheral arm of chromosome 5 and containing the SSI locus in chicory (upper part) and the corresponding region (~1.8 M) in endive located on the peripheral arm of chromosome 4 (bottom part). Along with all the predicted genes available from Fan et al., (Fan et al., 2022), we highlighted i) the AACC134 marker that delimits the abovementioned chromosomal windows, ii) the newly identified candidate gene (*MIK2*) in chicory and its putative orthologous in endive, and iii) a control gene (and its putative orthologous in endive) coding for a putative protein kinase, chosen as a control for a comparison between its polymorphism rate and that of the candidate gene.

In parallel, three out of seven markers (namely, AACC134, sw2H09.21 and B1311) were also successfully mapped against the endive genome (JAKOPN000000000), enabling us to establish the correspondence between chromosome 5 of chicory and chromosome 4 of endive. This corresponds to the observations made by Fan et al. based on the synteny analyses conducted between the two species (Fan et al., 2022). Additionally, complete collinearity was demonstrated in the order of these three markers within chromosome 4 of endive and chromosome 5 of chicory (**Figure 1A**).

According to the prediction made by Fan et al. (Fan et al., 2022), the ~4 M chromosomal window of chicory encompassing the SSI locus contains 139 genes, generically named "protein coding genes" (**Figure 1B**). The resulting proteins were aligned against the proteomes of Arabidopsis and lettuce, and 114 were successfully annotated (**Supplementary Table 1**). A gene encoding a protein with accession number KAI3736590.1 (1,038 aa) was considered a candidate gene of the SSI region. This gene (3,193 bp) was located between 2,811,314 bp and 2,814,506 bp and was orthologous to AT4G08850.1 (*A. thaliana*, AtMIK2 1076 aa) and XP_023768237.1 (*L. sativa*, LsMIK2 1036 aa). Both gene loci are annotated as *MDIS1 INTERACTING RECEPTOR LIKE KINASE 2* (*MIK2*), and the newly identified gene in chicory was therefore designated *CiMIK2*. In Arabidopsis, the MIK2 protein, along with MALE DISCOVERER 1 (MDIS1), MDIS2, and MIK1, forms a cell-surface male receptor complex (tetraheteromer) that is highly expressed on the pollen tube. Most interestingly, this receptor complex was found to perceive the female gametophyte-secreted peptide LURE1, also known as a female attractant (Wang et al., 2016). Based on the topology and protein domain prediction, CiMIK2 was characterized by an extracellular region of 676 aa containing 12 leucine-rich repeats (LRRs), a transmembrane region (TM, 10 aa) and a cytoplasmic region (in the C-terminal region, 323 aa) containing a kinase domain. This latter finding was particularly remarkable since, in the few species fully characterized for the self-incompatibility (SSI) system (e.g., *Brassica oleracea* and *B. campestris*), the female determinant of the SI locus is represented by a receptor kinase (SRK) (Tedder et al., 2011). In other plant species with SSIs, such as Convolvulaceae and Asteraceae, SRK-mediated self-recognition has been postulated (Hiscock et al., 2003).

The ortholog of *CiMIK2* was also detected in endive (*CeMIK2*, 1038 aa and a coding gene of 3200 bp) in the terminal region of chromosome 4, confirming the collinearity between this chromosomal arm and the terminal region of chromosome 5 in chicory (**Figure 1B**). From the nucleotide alignment between *CiMIK2* and *CeMIK2*, 116 positions out of a consensus sequence of 3200 bp were polymorphic, and 90 were nonsynonymous (90 amino acid changes out of 1038 positions, 8.67%). However, the distribution of the nonsynonymous positions throughout the sequence was nonuniform. In particular, the LRR-carrying extracellular region displayed a sequence identity between the two species equal to 89.35% and contained 72 of the 90 variable positions. In contrast, the cytoplasmic region was more conserved between chicory and endive, with a sequence identity of 95.80% and only 13 variable positions.

The control gene (KAI3736550.1, 496 aa and a coding gene of 1491 bp), located between 1,578,317 bp and 1,579,807 bp and, therefore, in the same chromosomal window carrying the SSI-locus (**Figure 1B**), was used to compare its polymorphism rate with that of *MIK2*. KAI3736550.1 was orthologous to AT1G28390.1 (*A. thaliana*, 475 aa) and XP_023768237.1 (*L. sativa*, 497 aa). Both genes, generically annotated as "protein kinase superfamily protein", do not seem to be involved in pollen–stigma recognition reactions based on the literature. KAI3736550.1 was chosen as a control not only for its proximity to the candidate gene but also for the presence of a kinase domain, which makes it structurally similar to *MIK2*. The orthologous protein of KAI3736550.1 was also identified in endive (KAI3513740.1), as expected, in the terminal region of chromosome 4 (**Figure 1B**). From the protein alignment between KAI3736550.1 and KAI3513740.1, 5/496 positions were polymorphic. The polymorphism between the two abovementioned protein sequences (1.01%) was therefore eight times lower than that observed when comparing the ciMIK2 and ceMIK2 proteins (8.67%).

### 3.2 Testing the polymorphism rate of *MIK2* in chicory and endive accessions

Loci governing SI are expected to experience negative frequency-dependent selection, a form of strong balancing selection where the relative fitness of a population increases as the frequency of each allelic variant decreases (Wright, 1939). In other words, low-frequency SI-related alleles benefit from a selective advantage over high-frequency alleles because they encounter their cognate allele only rarely, thereby enhancing cross-compatibility reactions. Consequently, in SI-related genes, substitutions affecting allelic specificity (i.e., nonsynonymous SNP, dN) are expected to enter a population more often than substitutions not affecting specificity (i.e., synonymous SNP, dS), leading to a positive dN/dS ratio (also known as ω) (Castric and Vekemans, 2007). This has been demonstrated, for example, for the S-locus receptor kinase (*SRK*) of *B. oleracea* and *B. rapa* (Sainudiin et al., 2005). In this study, we have addressed this issue by comparing the level of polymorphism of the newly identified candidate gene (*MIK2*) with that of the control gene, located in the same 4 M base window and encoding a putative protein kinase. For this, we successfully amplified and sequenced the full-length sequences encoding *MIK2* and the protein kinase in 15 samples belonging to *C. intybus* and *C. endivia*.

Regarding the control gene, by multiple alignment of the full sequences of the 15 samples along with those extrapolated from the chicory (*C. intybus* cultivar Grasslands Puna) and endive (*C. endivia* var. *crispum*) genomes, we detected 23 polymorphic sites over a full-length gene of 1,491 bp (98.46% of conserved nucleotides), with nearly 1 mutation every 64 nucleotides (**Supplementary Table 2**). Out of 23, 8 represented nonsynonymous mutations (highlighted as black blocks in **Figure 2A**) with a dN/dS of 0.53. This suggests that the gene is under negative (or purifying) selection.

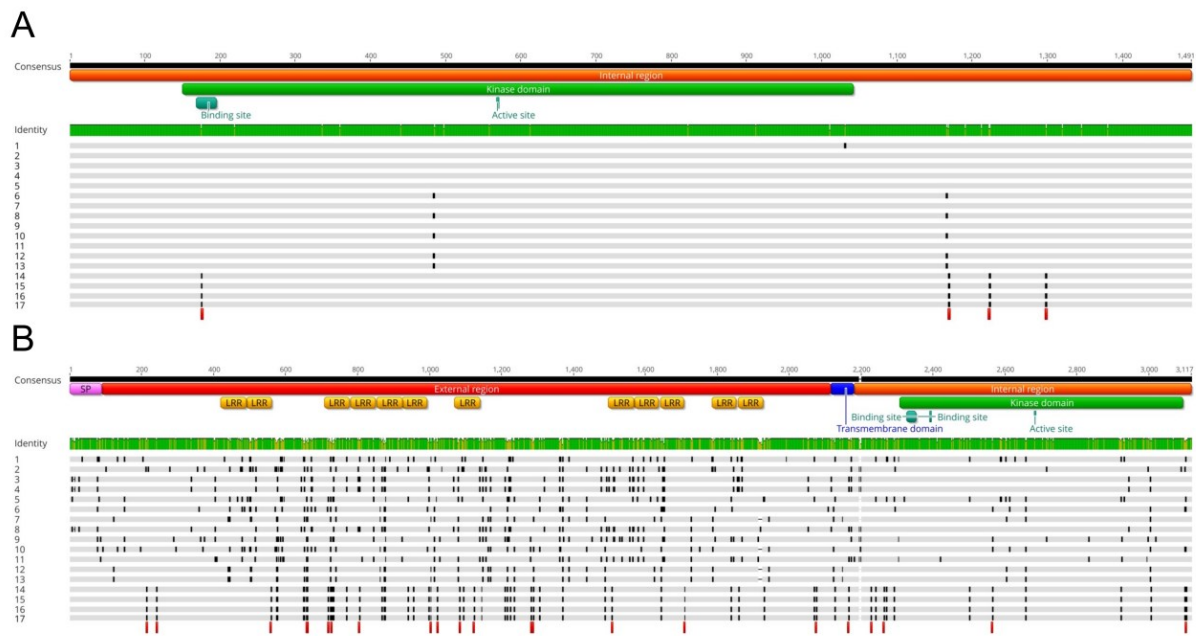

**Figure 2**. **(A)** Multiple alignment among the full CDSs of a putative kinase (used as control gene) sequenced in 12 chicory and 3 endive accessions. Along with the 15 newly obtained sequences, we also included the reference sequences extracted from the chicory and endive genomes (Fan et al., 2022). Sample 1 is the reference sequence extracted from the chicory genome (*C. intybus* cultivar Grasslands Puna). Samples from 2 to 13 are all Radicchio biotypes belonging to *C. intybus* botanical var. *foliosum* (from 2 to 5 Rosso di Chioggia, 6 Variegato di Chioggia, 7 and 8 Variegato di Castelfranco, 9 Rosa, 10 Verona Semilungo, 11 Treviso Tardivo, 12 and 13 Treviso Precoce). Samples 14 and 15 are accessions from *C. endivia* botanical var. *crispum* (the first one represents the reference sequence extracted from the endive genome), samples 16 and 17 are accessions from *C. endivia* botanical var. *latifolium.* Only nonsynonymous SNPs are shown in the alignment (black blocks), while red blocks indicate species-specific nonsynonymous SNPs (i.e., conserved within species but variable between chicory and endive. Cellular localization prediction, protein domains and functional sites of the resulting amino acid sequence are reported on the consensus sequence and were predicted by DeepTMHMM (https://dtu.biolib.com/DeepTMHMM) and PROSITE (Sigrist et al., 2010). **(B)** Multiple alignment among the full CDSs of *MIK2*, a candidate gene for the SI locus. Sample numbering, symbol legends and protein features are identical to those reported in Panel A.

When analyzing these sequences separately for each species, we found 9 polymorphic sites between the 14 chicory samples (3 nonsynonymous) and no polymorphism among the 4 endive sequences. Finally, in the comparison between chicory and endive samples, 11 polymorphic positions (4 nonsynonymous, highlighted with red blocks in **Figure 2A**) discriminated the two species (i.e., they were conserved within each species but variable between them).

Regarding the *MIK2* gene, multiple alignment of the CDSs (i.e., excluding the intron region of 76 nucleotides) revealed an astonishing scenario. The endive samples were 100% identical (no SNP, **Supplementary Table 3**), although they belonged to different botanical varieties (i.e., 2 var. *latifolium* and 2 var. *crispum*) and one of them (the one used for genome sequencing by Fan et al.) came from a geographical area (China) totally different from that of the other three (Italy).

In contrast, from the nucleotide alignment of the 14 chicory samples, we detected 387 polymorphic positions and 3 INDELs over a CDS consensus sequence of 3114 bp (87.48% of conserved nucleotides). This indicates 1 mutation every 8 nucleotides, eight times higher than that observed for the control gene. This result is even more relevant if we consider that, unlike endive, all the chicory biotypes analyzed belonged to the same botanical variety (*foliosum*). Enormous variability was found even within the same biotype. For example, Rosso di Chioggia 2 and Rosso di Chioggia 3 differed in 170 polymorphic positions. Similarly, Variegato di Castelfranco 2 and Variegato di Castelfranco 3 were found to be distinguishable for 141 SNPs. None of the 387 SNPs were found to produce nonsense mutations, and the INDEL lengths were always multiples of 3 bp (no frameshift). However, the number of nonsynonymous SNPs (265, black blocks in **Figure 2B**) was more than double that of the synonymous counterpart (122). The dN/dS ratio of 2.17 strongly suggests the possibility that the gene is under positive selection, as widely demonstrated for the SI-related genes of several species (Donia et al., 2016; Claessen et al., 2019; Azibi et al., 2020).

Additionally, as already observed in the preliminary analysis, the distribution of the nonsynonymous positions throughout the nucleotide sequence was uneven. The extracellular region (rich in LRRs) and the TM domain displayed the highest mutation rates (14.20 and 15.87 SNPs every 100 bases, respectively) and the highest dN/dS ratios (2.63 and 4.00, respectively). In species characterized by sporophytic or gametophytic SI, such as *Brassica* spp. (Sato et al., 2002), *Raphanus sativus* (Okamoto et al., 2004) and *Arabidopsis lyrata* (Miege et al., 2001), genes encoding components of the S-locus have been shown to possess regions of extreme sequence polymorphism, known as hypervariable (HV) domains. For example, in *Brassica* spp., the majority of sequence variation between SRKs lies within the extracellular domain. This HV region represents the receptor domain, which, by recognizing the male determinant, allows the stigma to discriminate between "self" and "nonself" pollen. The HV regions are therefore thought to be responsible for S-locus specificity (Ma et al., 2016). By functional and structural analogy, the extracellular part of ciMIK2 could represent one of the HV regions of the S-locus in chicory and the receptor region of the female determinant. In contrast, the cytoplasmic region (containing the kinase domain) was more conserved (dN/dS ratio of 1,08 and mutation rate of 8.76 SNPs every 100 bases).

It should be noted that in the comparison between chicory and endive samples, 25 polymorphic positions (of which 23 nonsynonymous positions are highlighted as red blocks in **Figure 2B**) discriminated the two species (i.e., they were totally conserved within each species but variable between them). Of these, 19 (including 18 nonsynonymous) were located in the extracellular region. The molecular mechanism by which two phylogenetically related and interfertile species differentiated to such an extent, such that endive evolved strictly autogamous (self-compatible) while chicory evolved strictly allogamous (self-incompatible), is still completely unclear. However, it is highly probable that the S-locus evolved differently in the two species, giving rise to contrasting sexual behaviors. The results observed from the comparison between the *ciMIK2* and *ceMIK2* sequences are an outstanding starting point, but they are not sufficient to support the hypothesis that the 23 species-specific nonsynonymous SNPs are actually responsible for their mode of reproduction.

Further studies are needed to understand whether the specific amino acid changes (especially the 18 identified in the receptor region) observed and conserved in the endive samples are actually responsible for alterations in protein folding and, possibly, in the receptor-male determinant interaction. The full sequence conservation observed between samples of different botanical varieties (*latifolium* and *crispum*) and from different geographical areas (China and Italy), along with the lack of nonsense mutations would suggest that *MIK2* retains its function in endive.

Considering the importance of the promoter in gene transcription regulation, we further investigated the first 500 bases upstream of the ATG codon in a subset of three chicory and three endive samples to identify species-specific polymorphisms that could explain any possible transcription change in the *MIK2* gene between the two species. Similar to what was observed for the coding sequence, we did not find any SNP (100% sequence identity) discriminating the promoter sequences of the three endive samples. In contrast, by comparing three chicory samples, we detected as many as 44 SNPs and 6 INDELs (**Supplementary Figure 1**).

Finally, in the comparison between species, 13 SNPs and 3 INDELs discriminated the two species (i.e., they were totally conserved within each species but variable between them). Most interestingly, the longest INDEL (i.e., a 10 bp INDEL) was located in one of the three CCAATBOX1 elements predicted through enriched motif screening. CCAAT box regions are universally known to be essential for gene expression in eukaryotic cells, contributing to transcription by recruiting a complex of nuclear factors: NF-YA, NF-YB, and NF-YC (Mantovani, 1999). Mutations in the CAAT box can lead to loss of NF-Y binding and, consequently, to decreased transcriptional activity (Zhong et al., 2023). Further investigations are needed to elucidate whether this specific promoter sequence variation could actually affect *MIK2* expression in the two species.

The next few years should be exciting for SI research in the *Cichorium* genus. Based on its chromosome location, the predicted protein structure, the role of its orthologs in Arabidopsis, and the impressive amount of nonsynonymous SNPs, we hypothesized that *MIK2* may represent the female determinant of the SSI locus in *C. intybus*. To establish that *MIK2* can act similarly to SRK in *Brassica*, the next major goal will be the functional characterization of this candidate gene to further corroborate its involvement in pollen–stigma recognition. In contrast, there are still no clues about the possible male counterpart (i.e., the male determinant). The full characterization of the S-locus will also enlighten the intricate phylogeny of the *Cichorium* genus and, in particular, the mechanism that led two very similar, closely related and interfertile crop plant species (i.e., endive and chicory) to adopt contrasting sexual reproduction strategies with significant consequences on the genetic structure and evolution dynamics of populations.

## 4 Conflict of Interest

The authors declare that the research was conducted in the absence of any commercial or financial relationships that could be construed as a potential conflict of interest.

## 5 Author Contributions

FP and GB: conceptualization. FP: methodology. SD: formal analysis. GM: data analysis. FP, GM, and GG: writing—original draft preparation. FP, SD, GM, GG, AV, SF, FS, ML, GB: writing—review and editing. FP and GB: supervision and project administration. GB: funding acquisition. All authors have read and agreed to the published version of the manuscript.

## 6 Funding

This study was performed within the Agritech National Research Center and received funding from the European Union Next-Generation EU (Piano Nazionale di Ripresa e Resilienza (PNRR)—Missione 4 Componente 2, Investimento 1.4—D.D. 1032 17/06/2022, CN00000022. Our study represents an original paper related to Spoke 1 "Plant, animal and microbial genetic resources and adaptation to climate changes". In particular, it is a baseline for the fulfilment of milestones within Task 1.3.5 titled "Genome-wide strategies for fast-forward molecular breeding aimed at the assessment of genetic distinctiveness, uniformity and stability (DUS) and identity of pre-commercial varieties". This manuscript reflects only the authors' views and opinions, and neither the European Union nor the European Commission can be considered responsible for them.

## 7 Data Availability Statement

The original contributions presented in the study are included in the article, in the Supplementary Materials and in GenBank. Further inquiries can be directed to the corresponding author.

## 8 Supplementary Material

The Supplementary Material for this article can be found at: